# Large-area fabrication of low- and high-spatial-frequency laser-induced periodic surface structures on carbon fibers

Clemens Kunz[1], Tobias N. Büttner[1], Björn Naumann[1], Anne V. Boehm[1], Enrico Gnecco[1], Jörn Bonse[2], Christof Neumann[3], Andrey Turchanin[3,4], Frank A. Müller[1,4], Stephan Gräf[1]*

[1]Otto Schott Institute of Materials Research (OSIM), Friedrich Schiller University Jena, Löbdergraben 32, 07743 Jena, Germany

[2]Bundesanstalt für Materialforschung und -prüfung (BAM), Unter den Eichen 87, 12205 Berlin, Germany

[3]Institute of Physical Chemistry, Friedrich Schiller University Jena, Lessingstrasse 10, 07743 Jena, Germany

[4]Center for Energy and Environmental Chemistry Jena (CEEC), Philosophenweg 7, 07743 Jena, Germany

*Abstract*

The formation and properties of laser-induced periodic surface structures (LIPSS) were investigated on carbon fibers under irradiation of fs-laser pulses characterized by a pulse duration $\tau = 300$ fs and a laser wavelength $\lambda = 1025$ nm. The LIPSS were fabricated in an air environment at normal incidence with different values of the laser peak fluence and number of pulses per spot. The morphology of the generated structures was characterized by using scanning electron microscopy, atomic force microscopy and Fast-Fourier transform analyses. Moreover, the material structure and the surface chemistry of the carbon fibers before and after laser irradiation was analyzed by micro Raman spectroscopy and X-ray photoelectron spectroscopy. Large areas in the $cm^2$ range of carbon fiber arrangements were successfully processed with homogenously distributed high- and low-spatial frequency LIPSS. Beyond

*Corresponding author. Tel: +49 (3641) 9-47754. E-Mail: Stephan.Graef@uni-jena.de (Stephan Gräf)

those distinct nanostructures, hybrid structures were realized for the very first time by a superposition of both types of LIPSS in a two-step process. The findings facilitate the fabrication of tailored LIPSS-based surface structures on carbon fibers that could be of particular interest for e.g. fiber reinforced polymers and concretes.

## 1.    Introduction

Carbon fibers are commercially used to reinforce polymers (carbon fiber reinforced polymers, CFRP) and concrete (engineered cementitious composites, ECC) [1-4]. The outstanding mechanical properties of single carbon fibers are a result of the specific structure consisting of graphitic and turbostratic carbon with a crystal structure very similar to graphite [5, 6]. Consequently, the strong covalent bonds between the carbon atoms of the graphene layers result in a high tensile strength in fiber direction. This anisotropic behavior also manifests in other fiber properties like crystallinity or electrical conductivity [1]. Regarding CFRP, the interface between carbon fibers and surrounding polymer matrix is in the focus of research activities aiming to increase fiber-matrix bonding strength and thereby to enhance the mechanical properties of the composite material. For this purpose, carbon fiber surfaces were modified by plasma oxidation, chemical and electrolytic etching as well as chemical vapor deposition [7-10]. It was shown that the bonding strength is improved by increasing the surface roughness and the resulting interface area between fiber and polymer matrix [11-15].

Concerning the engineering of surfaces with tailored functional properties, ultra-short pulsed laser processing gained rapidly increasing attention in the past decades. In this context, the fabrication of laser-induced periodic surfaces structures (LIPSS) in a single-step, direct-writing process emerged as a flexible and versatile technique [16]. LIPSS have been demonstrated as a universal phenomenon on almost all types of materials [17] providing outstanding properties of the laser-structured surface such as wettability, optical performance, bioactivity, and tribology [18, 19]. In particular, mimicking structures and functional principles provided by nature is an intensively studied field of current scientific research [20]. As a main advantage, the large variety of influencing parameters including laser wavelength $\lambda$, number of pulses $N$, pulse duration $\tau$, laser peak fluence $F$, angle of incidence $\theta$, and beam polarization, allows to control the specific properties of LIPSS such as their alignment and spatial period. Regarding the spatial period $\Lambda$, LIPSS are classified into low-spatial frequency LIPSS (LSFL) and high-spatial frequency LIPSS (HSFL). LSFL are characterized by a period $\Lambda$ close to the initial laser wavelength $\lambda$ for strong absorbing materials (metals, semiconductors) and close to $\lambda/n$ for dielectrics, where $n$ refers to the refractive index of the dielectric material [19, 21]. It is well-accepted, that the formation of LIPSS can be explained by spatial modulated intensity pattern resulting from interference of the incident laser radiation with surface electromagnetic waves that are generated by scattering at the rough surface [22]. This might include the excitation of surface plasmon polaritons (SPP) [22-24]. An alternative approach to explain LIPSS formation is given by a self-organization of the irradiated material via laser-induced thermal instabilities resulting in material redistribution [25]. HSFL with spatial periods much smaller than $\lambda$ are still controversially discussed in literature, and a deep understanding of the formation mechanism is still missing. Possible explanations include self-organization [26], chemical surface alterations [27] and second harmonic generation [28].

While the LIPSS formation on graphite was studied by several groups in dependence on different influencing parameters (beam polarization, laser peak fluence) on the flat surface of bulk materials [28-32], the formation of LIPSS on the surface of carbon fibers is less investigated [11, 33]. By performing single-spot experiments on strongly curved fiber surfaces, Sajzew et al. [33] demonstrated the formation of LSFL and HSFL within the Gaussian intensity distribution of the focal spot (diameter 50 µm) upon the irradiation of $N = 50$ linearly polarized fs-laser pulses with a peak fluence $F = 4$ J/cm$^2$. The study revealed, however, that the large number of laser pulses in combination with the utilized peak fluence lead to a remarkable ablation in the intense center of the focal spot and therefore to a damage of the fibers accompanied by a deterioration of their mechanical properties.

Based on the experimental findings of Sajzew at al., the objective of the present study is the homogenous manufacturing of HSFL and LSFL, respectively, on large areas of carbon fiber arrangements without damage by unidirectional scanning the focused laser beam over the sample surface. For this purpose, the nanostructure formation process was studied in dependence of the fs-laser peak fluence. The surface morphologies of the prepared samples were subsequently characterized by scanning electron microscopy (SEM) and atomic force microscopy (AFM). The material structure and the surface chemistry of the carbon fibers was studied before and after laser irradiation by micro Raman spectroscopy and X-ray photoelectron spectroscopy (XPS). Tensile strength measurements of single carbon fibers were performed in order to evaluate the impact of the fs-laser irradiation on the mechanical properties.

## 2. Materials and Methods

### 2.1 LIPSS formation on carbon fibers

The experimental setup is illustrated in Fig. 1. In order to fabricate large areas (5 x 5 cm$^2$) of LIPSS on carbon fiber surfaces, a diode pumped Yb:KYW thin disc fs-laser system (JenLas D2.fs, Jenoptik, Germany) with a central laser wavelength $\lambda = 1025$ nm and a pulse duration $\tau = 300$ fs was used. The emitted linearly polarized laser pulses are characterized by a repetition frequency $f_{rep} = 100$ kHz and pulse energies $E_{imp}$ up to 40 µJ. The laser beam was scanned with a galvanometer scanner (IntelliScan14, Scanlab, Germany) over the fiber surfaces. The scanner system is equipped with a 100 mm f-Theta objective (JENar, Jenoptik, Germany) that provides a spot diameter of the Gaussian beam $2w_f = (24 \pm 0.5)$ µm. In all experiments, the direction of laser scanning was perpendicular to the fiber axis and the direction of the beam polarization was parallel to the fiber axis (see also sketches in Figs. 2 and 3). For the structuring process, commercially available carbon fibers (Granoc XN-80-60s 6k 890 tex, Nippon Graphite Fiber Corporation, Japan) with a diameter of about 10 µm that have been prepared from a mesoporous pitch precursor were clamped into a fiber holder and placed under the scanner system (Fig. 1). This type of fixation is required to ensure a constant focal position of the flexible fibers and to prevent the deposition of vaporized components of an underlying substrate material. Moreover, it allows structuring almost the entire fiber surface in a two-step process based on the turning of the holder by 180°.

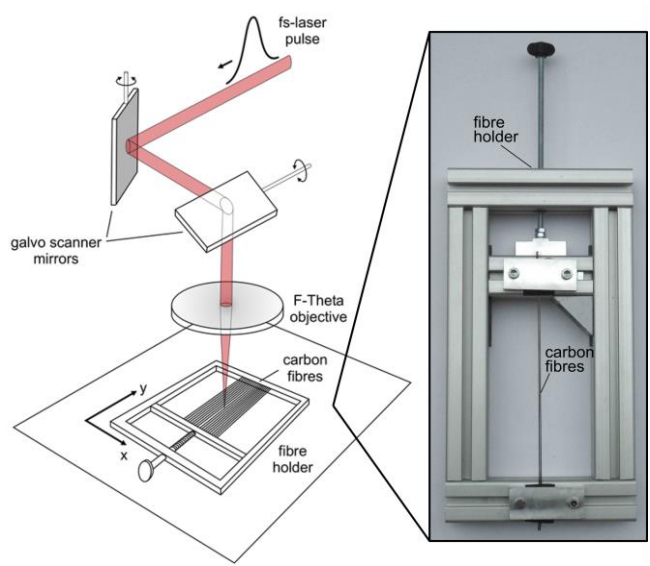

**Fig. 1: Experimental setup used for the fs-laser-induced formation of LIPSS on carbon fibers based on the specific positioning by means of a fiber holder.**

## 2.2 Characterization

The surface morphologies of the laser-processed samples were characterized by SEM (Sigma VP, Carl-Zeiss, Germany) at an accelerating voltage of 5 kV using the secondary electron detector. Height profiles of the laser-induced structures were measured by AFM (NanoWizard 4, JPK Instruments, Germany) working in the QI$^{TM}$ contact mode with a cantilever spring constant 0.14 N/m and a nominal tip radius of 2 nm. The LIPSS spatial periods were determined by Fast Fourier transform (FFT) of the SEM micrographs and confirmed by the AFM height profiles. In this context, the error bars indicate the width of distribution of the corresponding FFT operation. Optical images were taken with a digital camera in order to record the optical response (structural colors) of the fibers after laser processing.

Measurements of the tensile strength of single carbon fibers were performed using an universal testing machine (1445, Zwick Roell, Germany) with a 20 N load cell, a gauge length of 30 mm, a crosshead speed of 0.5 mm/min and a preload of 0.01 N. Single carbon fibers were fixed on a paper frame as described in ASTM Standard D3379-75 using epoxy resin adhesive.

The fiber material structure before and after fs-laser irradiation was analyzed using micro Raman spectroscopy (Senterra, Bruker, USA) operated in the backscattering mode. For this purpose, measurements at $\lambda = 532$ nm were obtained with a radiation power of 2 mW, a 50x objective and a thermoelectrically cooled CCD detector. The spectral solution of the system is 2-3 cm$^{-1}$. Before the measurement, the Raman peak of single-crystalline silicon at 520.7 cm$^{-1}$ was used for peak shift calibration of the instrument.

The surface chemistry was investigated by XPS (Multiprobe UHV system, Scienta Omicron, Germany) using a monochromatic X-ray source (Al K$_\alpha$) and an electron analyzer (Argus) with an energy resolution of 0.6 eV.

## 3.    Results and Discussion

### 3.1    Formation of HSFL

Figure 2 shows SEM micrographs of the carbon fiber surfaces before laser structuring (Fig. 2a) and after unidirectional scanning of the fs-laser beam over the fiber surface using a fs-laser peak fluence of $F = 0.4$ J/cm$^2$ (Fig. 2b), $F = 0.5$ J/cm$^2$ (Fig. 2c), $F = 0.7$ J/cm$^2$ (Fig. 2d) and $F = 0.9$ J/cm$^2$ (Fig. 2e). The scanning velocity was $v = 0.23$ m/s and the hatch distance was $\Delta x = 2$ µm resulting in an effective number of $N = 89$ linearly polarized laser pulses that hit the focal spot area. As indicated in Fig. 2, the direction of scanning is aligned perpendicular to the fiber axis and the linear beam polarization is parallel to the fiber axis. It becomes evident that the surface of the non-irradiated fiber is characterized by a shallow groove structure aligned along the fiber axis, which results from the stretching of the fibers during the manufacturing process (Fig. 2a). An irradiation with $F = 0.4$ J/cm$^2$ leads to the selective formation of HSFL with a spatial period $\Lambda \approx (79 \pm 37)$ nm ($\approx 0.077 \cdot \lambda$) and an alignment perpendicular to the beam polarization (Fig. 2b).

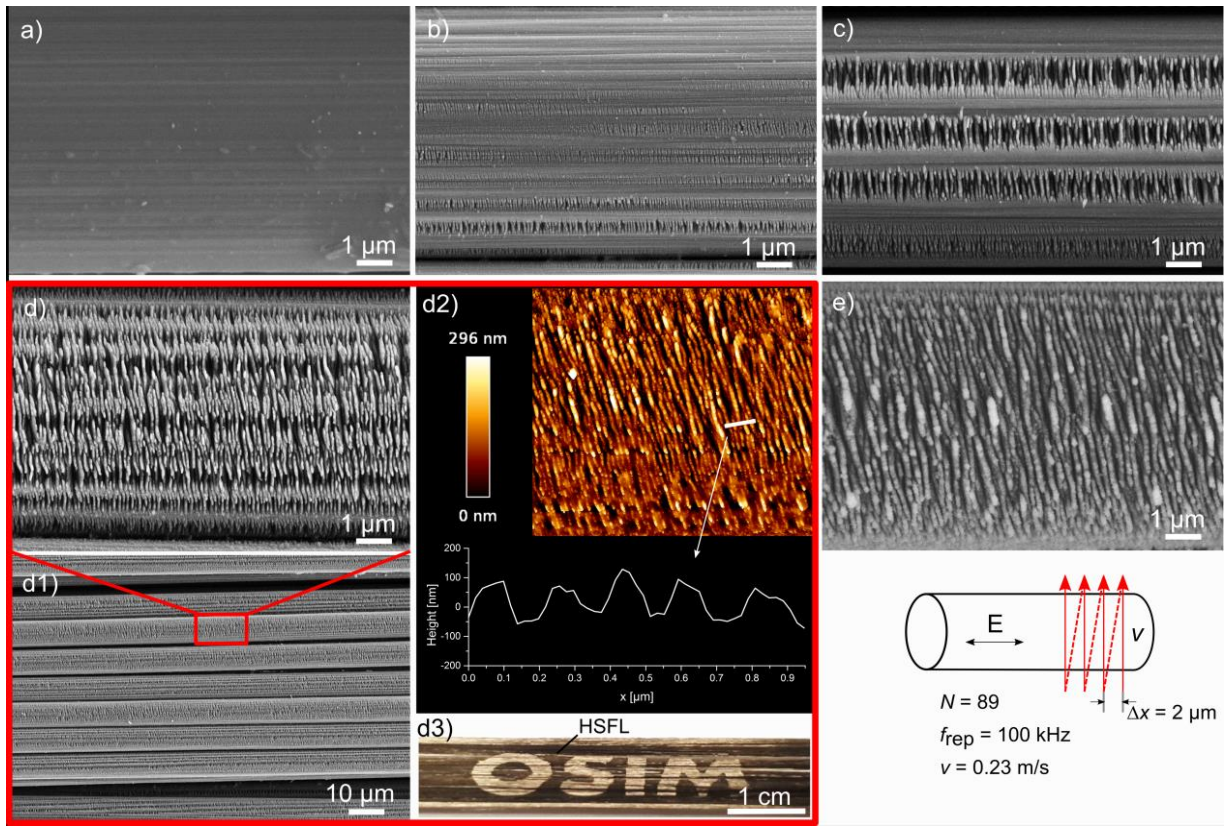

**Fig. 2: Fabrication of HSFL on carbon fibers by unidirectional scanning ($v = 0.23$ m/s, $\Delta x = 2$ μm, effective pulse number $N = 89$). SEM micrographs reveal the morphology of the fiber surface upon the irradiation with different fs-laser peak fluence: a) non-irradiated, b) $F = 0.4$ J/cm$^2$, c) $F = 0.5$ J/cm$^2$, d) $F = 0.7$ J/cm$^2$ and e) $F = 0.9$ J/cm$^2$. Using the parameters of the SEM micrograph in d) that resulted in a homogenous HSFL pattern d1) shows a SEM micrograph of a carbon fiber arrangement demonstrating the homogeneous large-area formation of HSFL, d2) shows an AFM micrograph of the HSFL with the corresponding surface height profile measured along the white line, and d3) provides a photograph of a carbon fiber arrangement indicating the optical response of HSFL that were processed outside the letters of the "OSIM" logo.**

As indicated in Fig. 2c, the increase of $F$ to 0.5 J/cm$^2$ results in an enhancement of the selective formation process. The generated HSFL are more pronounced and characterized by an increased spatial period $\Lambda \approx (89 \pm 72)$ nm ($\approx 0.087 \cdot \lambda$). The fiber surfaces irradiated with $F = 0.7$ J/cm$^2$ (Fig. 2d) exhibit a homogenous pattern across the entire curved fiber surface containing HSFL with a spatial period $\Lambda \approx (174 \pm 115)$ nm ($\approx 0.170 \cdot \lambda$). The selective formation of HSFL might be explained by the specific groove-like morphology of the initial fiber surface that leads to an inhomogeneous absorption of the laser radiation. Similar behavior has been reported by Nathala et al. for the formation of HSFL on flat titanium substrates with different surface roughness [34]. Contrary to mirror-like titanium surfaces where homogenously distributed HSFL are formed, the presence of initial tranches on the substrate surface resulted in the selective formation of HSFL, too. In accordance with our results, the increase of $F$ led to homogenously distributed HSFL independent of the initial roughness. It has to be noted, however, that the areas on the carbon fibers related to the onset of HSFL formation are still visible (Fig. 2d).

The AFM micrograph in Fig. 2d2 confirms the homogenous morphology of the fiber surface. Furthermore, the corresponding height profile of the HSFL obtained along the white line in the AFM micrograph reveals an almost sinusoidal surface pattern with an average height of about 150 nm and a spatial period similar to the value of the FFT of the SEM micrograph. Consequently, the depth-to-period-aspect-ratio is slightly larger than 1, hence the HSFL can be referred to deep-subwavelength ripples (HSFL-I) [19]. The increase of the laser peak fluence to $F = 0.9$ J/cm$^2$ (Fig. 2e) leads to a further increase of the spatial period, which was found to be $\Lambda \approx (261 \pm 157)$ nm ($\approx 0.255 \cdot \lambda$). However, the larger value of $F$ in combination with the large pulse number leads to a stronger ablation at the materials surface and therefore to a reduced fiber cross section at the area of normal incidence, which is undesirable due to deteriorated mechanical properties of the fibers. The determined increase of $\Lambda$ with increasing $F$ typical for HSFL was already observed on carbon fibers [33] and other materials [35].

Based on the findings concerning the formation of HSFL in dependence on $F$, we used $F = 0.7$ J/cm$^2$ (Fig. 2d) for fabrication of large areas with homogenously distributed HSFL. The corresponding SEM micrograph (Fig. 2d1) proves that the exact positioning of the fibers by means of the fiber holder and the well-defined adjustment of the processing parameters allow to fabricate large-area HSFL. The uniformity of the structures is also confirmed by the optical photograph of the fiber arrangement with a size of about (5 x 1) cm$^2$ (Fig. 2d3) that demonstrates the optical response of the HSFL covering regions outside the letters of the "OSIM" logo. The photograph reveals that the areas containing HSFL appear darker than the non-structured areas under illumination with white light. This can be explained based on the results reported by Calvani et al. concerning the optical properties of HSFL on fs-laser treated diamond [36]. Total transmission and reflection measurements using an integrating sphere demonstrated that HSFL with $\Lambda = (170 \pm 10)$ nm enhance light trapping so that absorbance strongly increases to a maximum of about 80% in a wide spectral range, in particular in the visible and IR spectral ranges. In these investigations, the increased absorbance was shown to be mainly correlated to a decreased reflectance of the materials surface [36]. Moreover, contributions by scattering effects at the sub-wavelength structures cannot be excluded here.

## 3.2 Formation of LSFL

Figure 3 shows a comparison of SEM micrographs obtained from a non-irradiated fiber surface (Fig. 3a) and from fiber surfaces structured with different values of $F$ between 1.6 J/cm$^2$ and 4.8 J/cm$^2$ (Fig. 3b-e). According to Sajzew et al. [33], the effective number of linearly polarized laser pulses per focal spot area was reduced to $N = 12$ in order to fabricate solely LSFL without mechanical damage of the fibers by excessive ablation. For this purpose, the scanning velocity was set to $v = 0.63$ m/s and the hatch distance to $\Delta x = 6$ µm. The SEM micrograph obtained from the carbon fiber surface upon the irradiation with the lowest value $F = 1.6$ J/cm$^2$ shows the first appearance of LSFL (Fig. 3b), which exhibit a relatively weak surface modulation with a spatial period $\Lambda = (872 \pm 62 \text{ nm})$. Analogue to HSFL and in good agreement to literature [29, 33], the orientation of LSFL is also perpendicular to the linear beam polarization. For $F < 1.6 \text{ J/cm}^2$ (not shown here), the possible transition between HSFL and LSFL was not observed because of the insufficient number of laser pulses.

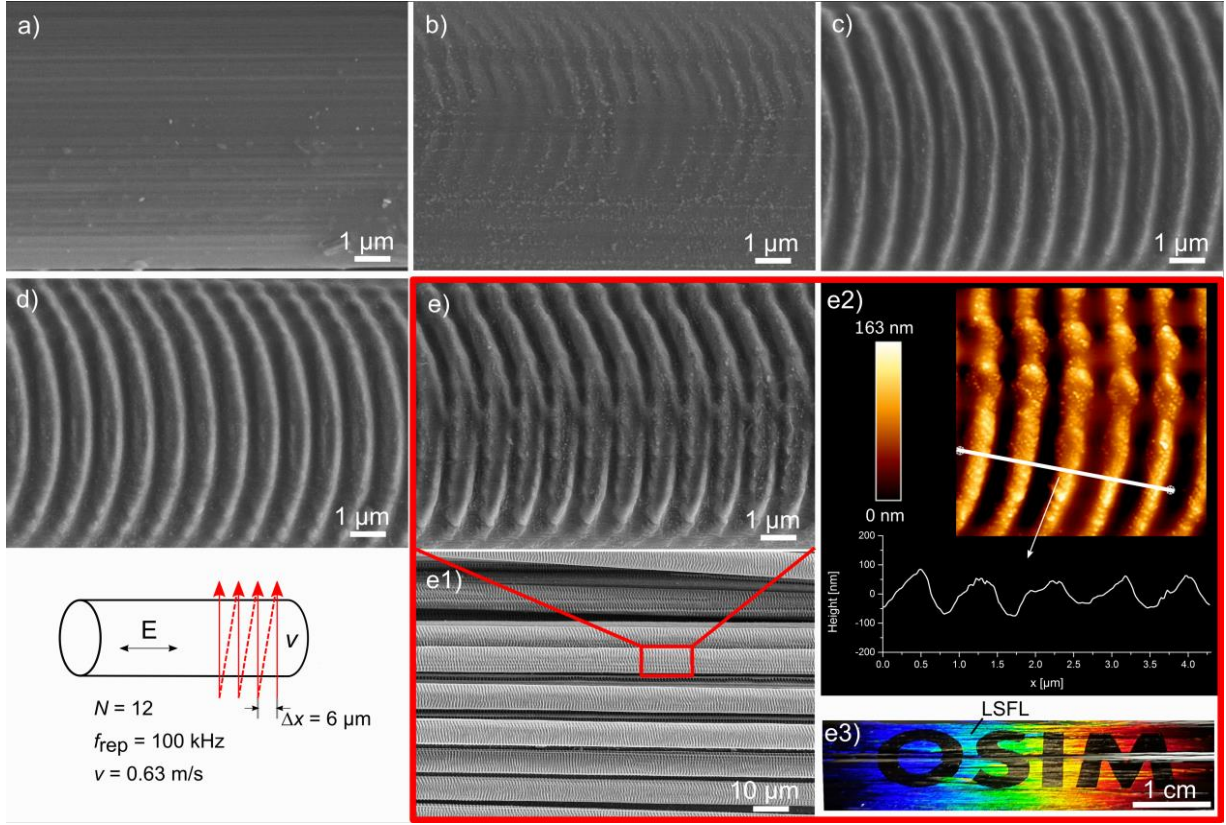

**Fig. 3: Fabrication of LSFL on carbon fibers by unidirectional scanning ($v = 0.63 \text{ m/s}$, $\Delta x = 6 \text{ µm}$, effective pulse number $N = 12$). SEM micrographs reveal the morphology of the fiber surface upon the irradiation with different fs-laser peak fluences: a) non-irradiated, b) $F = 1.6 \text{ J/cm}^2$, c) $F = 2.6 \text{ J/cm}^2$, d) $F = 3.5 \text{ J/cm}^2$ and e) $F = 4.8 \text{ J/cm}^2$. Using the parameters of the SEM micrograph in e) that resulted in a homogenous well-pronounced LSFL pattern e1) shows a SEM micrograph of a carbon fiber arrangement demonstrating the homogeneous large-area formation of LSFL, e2) shows an AFM micrograph of the LSFL with the corresponding surface height profile measured along the white line, and e3) provides a photograph of a carbon fiber arrangement indicating the optical response of LSFL.**

It becomes evident from the SEM micrographs that with increasing $F$ the LSFL become increasingly pronounced, whereby the spatial period $\Lambda$ remains almost unaffected by the laser peak fluence at $\Lambda = (902 \pm 92)$ nm for $F = 2.6$ J/cm$^2$ (Fig. 3c), $\Lambda = (942 \pm 77)$ nm for $F = 3.5$ J/cm$^2$ (Fig. 3d), and $\Lambda = (942 \pm 87)$ nm for $F = 4.8$ J/cm$^2$ (Fig. 3e). Thus, the corresponding center positions of the characteristic peaks of the FFT operations indicate spatial periods close to the initial laser wavelength in the range 0.88-0.92·$\lambda$. Feng et al. [29] reported similar results for the irradiation of highly oriented pyrolytic graphite in ambient air atmosphere. The spatial periods were in the range of $\Lambda = 712$ nm ($\approx 0.89·\lambda$) using 800 nm pulses with a pulse duration of $\tau = 50$ fs. Finally, cross-sections in the corresponding AFM micrograph of the surface structured with $F = 4.8$ J/cm$^2$ reveal an average height of about 150 nm (Fig. 3e2), which is quite similar to the height of the HSFL.

The homogeneity of the LSFL obtained from large-area scanning of carbon fiber arrangements with $F = 4.8$ J/cm$^2$, $v = 0.63$ m/s, $\Delta x = 6$ µm, and $N = 12$ is illustrated by the respective SEM micrograph in Fig. 3e1 and the optical photograph in Fig. 3e3. The latter illustrates the colorizing effect of the LSFL, which is related to varying structural colors observed at different viewing angles upon the illumination by white light. These colors originate from the diffraction behavior of the grating-like LSFL structures [37-41]. They were reported for the very first time by Vorobyev and Guo for fs-LSFL, who demonstrated that the colorizing of metal surfaces facilitates to control their respective optical properties and in particular their color [37]. In the present study, they were used to illustrate the capability large area structuring in the cm$^2$ range as well as the homogeneity of the grating structures.

### 3.3 Laser-induced hybrid structures

By combining the above results in a two-step structuring process, a combination of both types of LIPSS could be achieved. These hybrid structures were realized by the generation of well-pronounced LSFL ($F = 4.8$ J/cm$^2$, $N = 12$) in a first step and the subsequent superposition of HSFL ($F = 0.7$ J/cm$^2$, $N = 89$) in a second step. Figure 4 shows SEM micrographs (Fig. 4a) and AFM micrographs (Fig. 4b,c) of these structures.

It becomes evident that both, HSFL and LSFL, can be observed in the final surface morphology leading to a hybrid structure consisting of micro- and nano-sized surface structures. Note that literature concerning such a superposition of both LIPSS types is limited to metals [42-44].

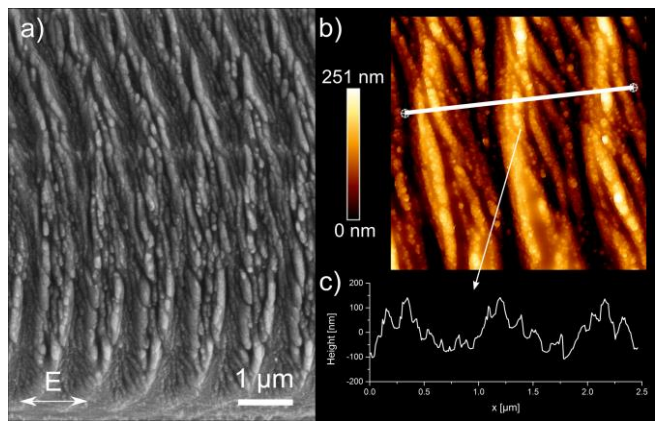

**Fig. 4: Generation of hybrid structures on carbon fibers by the superposition of LSFL ($F$ = 4.8 J/cm$^2$, $N$ = 12) and HSFL ($F$ = 0.7 J/cm$^2$, $N$ = 89) based on a two-step structuring process: a) SEM micrograph, b) AFM micrograph and c) corresponding surface height profile measured along the white line in b).**

Contrary to the hybrid structures illustrated in Fig. 4, however, the corresponding investigations revealed that LSFL and HSFL are aligned perpendicular to each other and that the HSFL can only be observed in the valleys of the grating-like surface structures. These novel hybrid structures built up by a superposition of HSFL and LSFL might be of potential interest for various applications in the fields of optics, wettability and biomaterials [18, 45, 46].

### 3.4 Characterization of the carbon fibers

### 3.4.1 Mechanical Properties

Tensile strength measurements were performed with single carbon fibers at a fixed gauge length of 30 mm before and after fs-laser irradiation with the highest laser peak fluence $F$ = 4.8 J/cm$^2$ and a pulse number of $N$ = 12. The resulting structures correspond to LSFL illustrated in Fig. 3e. Forty reference and irradiated samples, respectively, were tested in order to evaluate the impact of fs-laser structuring. The measurements revealed an average tensile strength of $\sigma_f = (2.5 \pm 0.7)$ GPa for the non-structured fibers and of $\sigma_f = (2.5 \pm 0.4)$ GPa for the fibers structured with LSFL. Consequently, a negative impact of the fs-laser irradiation on the mechanical properties of carbon fibers can be excluded. This is reasonable, as the depth of the LSFL of about 150 nm (Fig. 3e2) is negligibly small when compared to the initial fiber diameter of 10 µm, i.e. in the valleys of the LSFL pattern the fiber diameter is only reduced by 3% and the effective cross-sectional area by 5.9%, respectively. It has to be noted that the values measured for the single carbon fibers are below the tensile strength of $\sigma_f = 3.4$ GPa specified by the manufacturer. Huang and Young [47] demonstrated for different types of fibers that $\sigma_f$ strongly depends on the gauge length, i.e. $\sigma_f$ increases with decreasing gauge length towards the intrinsic strength of the fibers. This behavior can be explained by the presence of defects and flaws [47], whose amount depends on the gauge length as well.

### 3.4.2 Fiber structure and surface chemistry

The material structure of the carbon fibers before and after laser irradiation was studied by micro Raman spectroscopy (Fig. 5). The spectrum of the non-irradiated sample represents a typical high modulus pitch-based carbon fiber [47]. It is characterized by the characteristic $G$ peak at $\approx 1585$ cm$^{-1}$ resulting from the in-plane stretching mode of the graphite planes and the $D$ peak at $\approx 1355$ cm$^{-1}$ indicating disorder and/or defects in the graphene structure [48, 49]. Other peaks visible are those related to defect activated bands in sp$_2$ carbon materials ($D'$ at $\approx 1622$ cm$^{-1}$) and to turbostratic orientation of the graphene layers ($2D$ at $\approx 2700$ cm$^{-1}$) [48-50]. Moreover, a small peak termed as $D+D''$ arises at $\approx 2450$ cm$^{-1}$ [49]. The degree of disorder can be quantified by the ratio $I_D/I_G$ of the $D$ and $G$ peak intensities. For the non-irradiated fibers, $I_D/I_G$ was calculated to be 0.27, which indicates that the investigated carbon fibers exhibit a moderate defect density in the graphite layers. After the fabrication of HSFL with $F = 0.7$ J/cm$^2$ and $N = 89$ (Fig. 2d), the respective peaks broaden and $I_D/I_G$ increases to about 0.57 due to an increased *D* band intensity. Furthermore, the 2*D* peak intensity decreases and the *D* and *G* peak widths increase upon fs-laser irradiation. This behavior indicates an increased structural disorder in the graphite crystalline structure induced by the fs-laser irradiation [49, 51]. Similar behavior has been reported by Feng et al. [29] for LIPSS formation on highly-oriented pyrolytic graphite at ambient air atmosphere using 50 fs laser pulses in the fluence range 0.2-2 J/cm$^2$. It becomes evident from Fig. 5, that both Raman spectra obtained from the structured fibers are almost similar, although the fabrication of LSFL required a larger laser peak fluence (*F* = 4.8 J/cm$^2$). This might be explained by the much lower number of laser pulses utilized in the case of LSFL (*N* = 12).

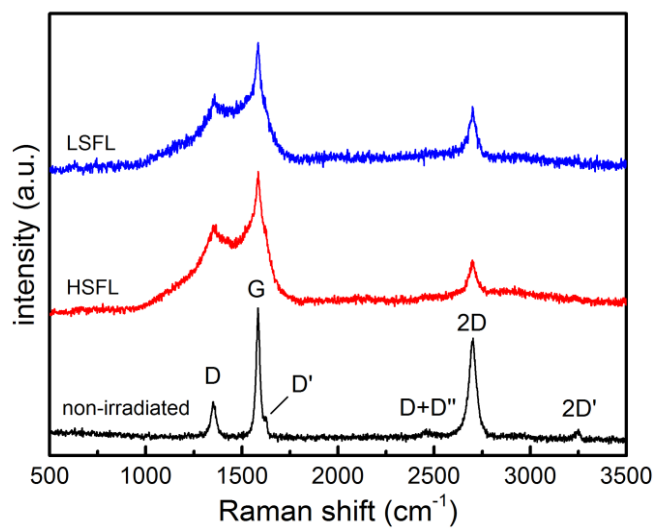

**Fig. 5: Raman spectra of carbon fibers: a) non-irradiated fiber, b) fiber covered with HSFL (*F* = 0.7 J/cm$^2$, *N* = 89) and c) fiber structured with LSFL (*F* = 4.8 J/cm$^2$, *N* = 12). The spectra are vertically shifted for better visibility.**

For an analysis of the chemical composition of the carbon fibers before and after fs-laser irradiation, XPS was performed (Fig. 6). The C1s signal of the non-irradiated fiber consists of a main peak at a binding energy (BE) of 285.0 eV (C-C sp$_3$, black) accompanied by a shoulder at 284.3 eV (C-C sp$_2$, red). In contrast to the results obtained by Raman spectroscopy, where the sharp *G* and 2*D* bands indicate the prevailing presence of sp$_2$ type of carbon, the intensity of the peaks assigned to sp$_3$ type of carbon dominates the C1s signal in the XPS spectrum (Tab. 1). This difference is due to different sampling depths of XPS and Raman spectroscopy. While Raman spectroscopy has a sampling depth of ~ 1 µm, XPS is a very surface sensitive technique. With an inelastic mean free path for C1s electrons in our experiment of $\lambda = 32$ Å, the sampling depth corresponds to $3 \cdot \lambda = 10$ nm [52]. As the non-irradiated fiber is typically coated with a sizing agent of a ~ 100 nm thick epoxy polymer [53, 54], only this mostly $sp_3$ type of carbon containing layer is probed. The carbon species at a BE of 286.6 eV (blue), which can be mainly attributed to C-O groups is also in agreement with typical XP spectra of sized fibers [55]. The measured O1s signal (Fig. 6b) confirms the results as a pronounced peak at a BE of 533.3 eV (C-O, red) is found with a shoulder attributing to a small amount of C=O double bonds within the sizing agent (531.8 eV green).

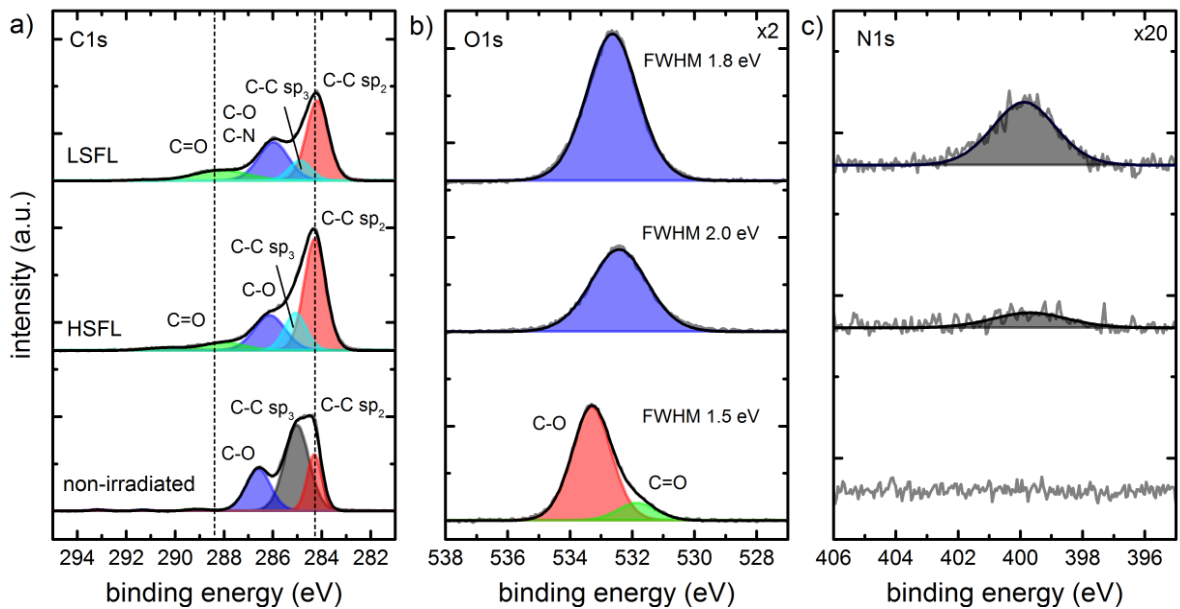

**Fig. 6: XPS spectra of non-irradiated carbon fibers in comparison to carbon fibers structured with HSFL ($F = 0.7$ J/cm², $N = 89$) and LSFL ($F = 4.8$ J/cm², $N = 12$): a) C1s peaks, b) O1s peaks and c) N1s peaks. Intensities of the O1s and N1s spectra were multiplied by 2 and 20, respectively.**

After fs-laser irradiation, the characteristics of the spectra change. The C1s peak, both of HSFL and LSFL shifts towards lower binding energies (284.2 eV, red) indicating the presence of mainly $sp_2$ type of carbon as expected for the irradiated fiber after removal of the sizing agent by the laser beam. The measured amount of $sp_3$ type of carbon is reduced in agreement with the introduced disorder in the sample found by the rise of a *D*-peak in the Raman spectra (Fig. 5). Therefore, this carbon species has a different origin than in the non-irradiated fibers and is marked in a different color (Fig. 6a, light blue). Due to the fs-laser irradiation a new carbon species marked by the dashed vertical line arises in the spectra at a binding energy of ~ 288 eV and can be attributed to C=O bonds. As the surface structuring was performed at ambient conditions, the increased amount of this species is related to the higher laser fluences. The creation of additional C=O bonds is supported by the broadening of the FWHM of the O1s peak from 1.5 eV for the non-irradiated fiber towards 1.8-2 eV after irradiation (Fig. 6b). Due to the high laser fluences most likely a laser-induced plasma is formed in the interaction zone leading not only to the formation of additional carbon-oxygen bonds but also to the incorporation of nitrogen from the air to the fiber. This assumption is supported by the N1s spectra measured before and after fs-laser irradiation, where a peak at a BE of 399.8 eV emerges (Fig. 6c), being indicative of C-N or C-N-H bonds. The measured amount of nitrogen increases with increasing laser fluence from 0.6 at% (HSFL) to 2.0 at% (LSFL).

**Tab. 1: Peak assignments, binding energy and the percent contribution of the peaks for the non-irradiated fiber as well as for HSFL ($F = 0.7$ J/cm$^2$, $N = 89$) and LSFL ($F = 4.8$ J/cm$^2$, $N = 12$).**

| Type | | C1s | | | | | O1s | | N1s |
|---|---|---|---|---|---|---|---|---|---|
| | | C-C (sp$_2$) | C-C (sp$_3$) | C-O | C=O | Shake ups | O=C | O-C | N-C |
| LSFL | BE (eV) | 284.2 | 284.9 | 286.0 | 288.1 | >289 | 532.6 | | 399.9 |
| | Area (%) | 45 | 11 | 29 | 14 | 1 | 100 | | 100 |
| HSFL | BE (eV) | 284.3 | 285.1 | 286.1 | 288.0 | >289 | 532.8 | | 399.7 |
| | Area (%) | 50 | 17 | 23 | 7 | 3 | 100 | | 100 |
| non-irradiated | BE (eV) | 284.3 | 285.0 | 286.6 | - | >289 | 531.8 | 533.3 | - |
| | Area (%) | 20 | 52 | 26 | - | 2 | 13 | 87 | - |

### 3.5 Theoretical Analysis of LIPSS Formation

Figure 7 shows the spatial periods $\Lambda$ that were realized on the carbon fiber surfaces in dependence on the fs-laser peak fluence $F$ and pulse number $N$. The specific values were obtained from Fast-Fourier transform (FFT) of the corresponding SEM micrographs (Figs. 2 and 3). Taking into account the width of the LIPSS features in the reciprocal space, as obtained by FFT, which is illustrated by the vertical bars in Fig. 7, spatial periods $\Lambda$ in the ranges of 17 to 400 nm and 810 to 1029 nm can be observed for HSFL and LSFL, respectively.

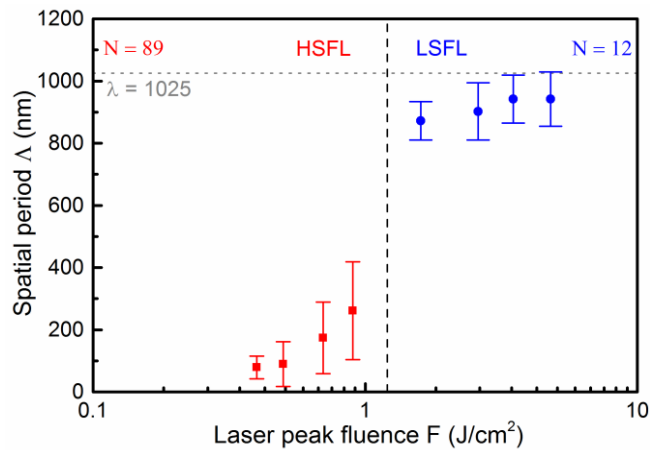

**Fig. 7: Spatial periods $\Lambda$ of HSFL and LSFL fabricated on carbon fibers in dependence on the laser peak fluence $F$ and number of pulses $N$. The vertical bars indicate the width of the LIPSS features in the reciprocal space, as obtained by FFT.**

Theoretical analysis of LIPSS formation was performed using the widely accepted theory of Sipe et al. [22], which is based on the interference of the incident laser beam with surface electromagnetic waves that are generated by scattering at the rough surface. This theory predicts possible LIPSS wave vectors $\boldsymbol{k}$ of the surface (with $|\boldsymbol{k}| = 2\pi/\Lambda$), i.e. it allows to calculate the spatial periods and the orientation of LIPSS using the optical properties of the material as input data. The predicted wave vectors depend on the laser parameters including the angle of incidence $\theta$, the polarization direction, and the wave vector $k_L = 2\pi/\lambda$ of the incident laser radiation (which has a component $k_i$ in the surface plane), as well as on the sample properties such as the bulk dielectric constant $\varepsilon^*$ and the surface roughness parameters. The theory provides the so-called "efficacy factor" $\eta$ that determines the efficacy of a surface to absorb energy at the wave vector $\boldsymbol{k}$. The scalar function $\eta$ can exhibit sharp peaks at certain $k$ values, which determine the spatial periods of LIPSS. For the analytical calculation of $\eta$ as a function of the normalized LIPSS wave vector components $\kappa_x$ and $\kappa_y$ ($|\kappa| = \lambda/\Lambda$), we used the simplified set of equations derived from the original Sipe theory by Bonse et al. [56]. It has to be noted that the calculation of $\eta$ refers to flat substrate materials. Due to the observed graphitic structure of the non-irradiated carbon fibers (Fig. 5) and the relatively small absorption length ($\approx 12$ nm at the utilized laser wavelength) [57], we used the complex refractive index of bulk graphite for the calculations. Generally, the optical properties of non-excited graphite are determined by the complex refractive index $n^* = n + \mathrm{i}k = 3.11 + \mathrm{i} \cdot 1.94$) [58], where $n$ is the refractive index and $k$ is the extinction coefficient. Graphite exhibits a semi-conductor like behavior because of the predominating interband transitions [32]. Consequently, the transient change of the optical properties due to the excitation of quasi-free electrons can be described by a Drude model [59], which provides the dielectric function of the laser-excited material $\varepsilon^* = n^{*2}$ by adding the additional Drude term $\Delta\varepsilon_D$ to the dielectric function $\varepsilon = \varepsilon_r + \mathrm{i}\varepsilon_i$ of the non-excited material:

$$\varepsilon^* = \varepsilon + \Delta\varepsilon_D = \varepsilon - \frac{e^2 \cdot N_e}{\varepsilon_0 \cdot m_{opt} \cdot m_e \cdot \omega^2}\left(1 + \frac{i}{\omega \cdot \tau_D}\right) \qquad (1)$$

Here, $e$ represents the electron charge, $N_e$ is the laser-induced electron density in the conduction band of the solid, $m_e$ is the electron mass, $\varepsilon_0$ is the vacuum dielectric permittivity and $\omega$ the laser angular frequency. For fs-laser-excited graphite, an optical effective mass of the carriers $m_{opt} = 0.024$ and a Drude damping time $\tau_D = 0.8$ fs have been chosen according to the work of Golosov and co-workers [32]. Figure 8 illustrates the carrier density dependence of the optical properties of graphite calculated by Drude model (Eq. 1). The graph reveals that for $N_e > 1 \cdot 10^{20}$ cm$^{-3}$ the optical constants $n$ and $k$ are strongly affected by $N_e$. As a consequence, the reflectivity of normal incident light at the laser-excited graphite surface rapidly increases above $N_e \approx 2 \cdot 10^{20}$ cm$^{-3}$, where the material evolves from its semiconductive behavior into a metal-like, high-reflective state characterized by Re[$\varepsilon^*$] < 0.

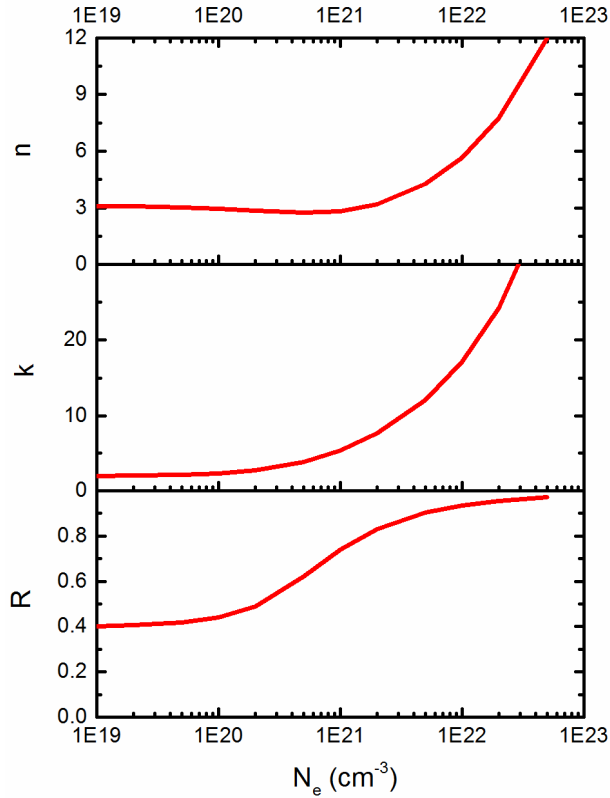

**Fig. 8: Optical properties of fs-laser irradiated graphite calculated in dependence of the laser-induced electron density $N_e$ using the parameters $\lambda = 1025$ nm, $m_{opt} = 0.024$, and $\tau_D = 0.8$ fs; $R$: reflectivity at normal incidence, $n$: refractive index, $k$: extinction coefficient.**

Fig. 9a-d shows the resulting two-dimensional grey-scale images of $\eta$ calculated for laser-irradiated graphite as a function of the normalized LIPSS wave vector components $\kappa_x$, $\kappa_y$ ($|\kappa| = \lambda/\Lambda$) for different degrees of material excitation ($N_e$). For the non-excited material ($N_e = 0$, $\varepsilon^* = 5.91 + \mathrm{i} \cdot 12.07$), the grey-scale image (Fig. 9a) exhibits two sickle-shaped regions, where $\eta$ is increased. Sharp $\eta$-maxima can be obtained at $\lambda/\Lambda_y = \pm 1.027$. They can be associated to LSFL with spatial periods around $\Lambda \approx 997$ nm and with an orientation perpendicular to the beam polarization (indicated by horizontal arrows in Fig. 9a). Deviations from the experimental

LSFL spatial period $\Lambda = (872 \pm 62)$ nm observed at the lowest fluence $F = 1.6$ J/cm$^2$ can be explained by the fact that the theory of Sipe refers to a single-pulse interaction process. In this context, it was shown for other semiconductors and metals that $\Lambda$ decreases with increasing pulse number $N$ [17, 19]. Moreover, inter-pulse effects such as grating-assisted coupling are not considered in the Sipe theory although they strongly influence the formation process [24]. Under consideration of the transient optical properties, however, Fig. 9b-d illustrates that the increase of $N_e$ leads to a modification of the sickle-like shaped features, which also corresponds to the cross-sections of the $\eta$-maps along the positive $\kappa_y$-direction at $\kappa_x = 0$ (Fig. 9e). According to Sipe et al., LIPSS can be expected where $\eta$ exhibits a sharp maxima or minima [22]. The graphs calculated for different values of $N_e$ demonstrate that with increasing $N_e$ the position of the $\eta$-maximum is shifted to somewhat smaller $\kappa_y$-values, i.e. towards larger spatial periods. The shift of the LSFL features is in line with the slight increase of the spatial periods with increasing laser peak fluence $F$ (Fig. 7). With increasing $F$, more electrons are excited into the conduction band, which results in an increase of $N_e$ and consequently of $\Lambda$. At an electron density of $N_e \approx 5 \cdot 10^{20}$ cm$^{-3}$, the LSFL feature exhibits a very sharp sickle-shaped contour indicating the excitation of SPP. The SPP activity of the laser-excited material at that carrier density is supported by the value of the dielectric function which accounts to $\varepsilon^* = -7.5 + \mathrm{i} \cdot 21.17$. The latter fulfils the condition $\mathrm{Re}[\varepsilon^*] < -1$ usually associated with the excitation of SPP [60, 61]. At carrier densities exceeding $1 \cdot 10^{21}$ cm$^{-3}$, the LSFL features vanish in the $\eta$-map. This behavior is similar to the case of fs-laser excited silicon [23] and is consistent with the general observation that an upper fluence limit exists for the observation of LSFL [19, 23]. Note that the HSFL are not properly described by Sipes theory [19]. Hence, numerical methods [62, 63] may be applied and adapted to the carbon material here, which are beyond the scope of this work.

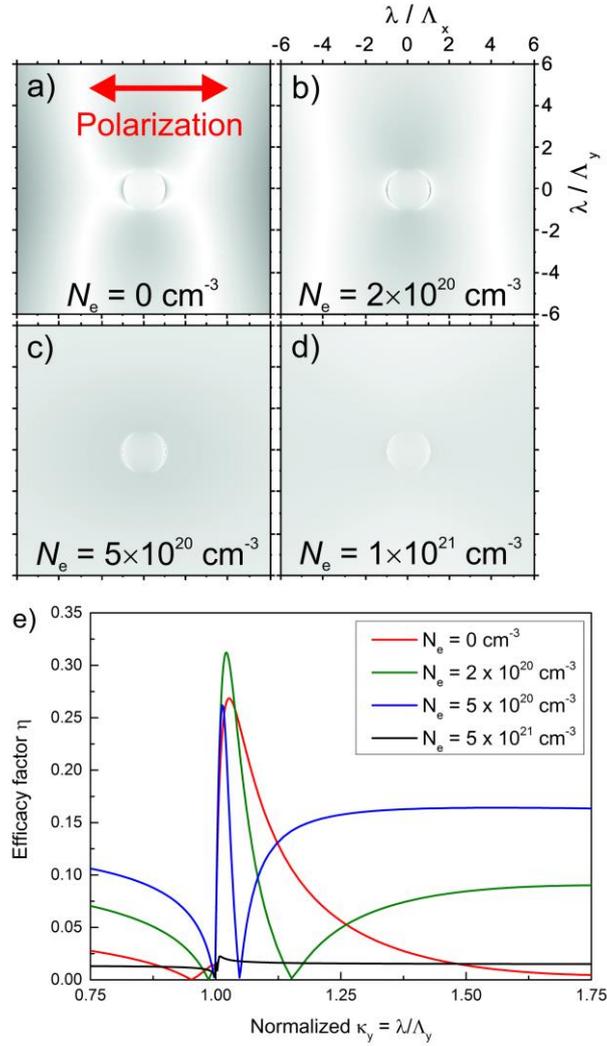

**Fig. 9: Efficacy factor $\eta$ calculated for graphite as a function of the normalized LIPSS wave vectors $\kappa_x$, $\kappa_y$ with $\lambda = 1025$ nm and $\theta = 0°$: a-d) Two-dimensional grey-scale images of $\eta$ and e) cross-sections of the $\eta$-maps along the positive $\kappa_y$-direction at $\kappa_x = 0$ in dependence of the electron density $N_e$ a) $N_e = 0$ ($n = 3.110$, $k = 1.940$), b) $N_e = 2 \cdot 10^{20}$ cm$^{-3}$ ($n = 2.852$, $k = 2.754$), c) $N_e = 5 \cdot 10^{20}$ cm$^{-3}$ ($n = 2.735$, $k = 3.870$), d) $N_e = 5 \cdot 10^{21}$ cm$^{-3}$ ($n = 4.263$, $k = 12.095$). In a-d), the values of $\eta$ are encoded in a common linear grey scale with dark colors representing larger values.**

## 4. Conclusion

The formation of laser-induced periodic surface structures on carbon fibers was studied in dependence of the laser peak fluence and the number of laser pulses. It was show that large areas of carbon fiber arrangements can be structured with HSFL and LSFL by scanning the fs-laser beam over the substrate surface. Beyond, novel hybrid structures were realized for the very first time by superimposing both LIPSS types. Tensile tests confirmed that the tensile strength remained unaffected by the formation of LSFL. Micro Raman and XPS analysis revealed a moderate increase of the structural disorder of the carbon fibers after LIPSS formation. Furthermore, the formation of additional carbon-oxygen bonds and the implementation of nitrogen at the fiber surface was shown. Considering the observed results and the beneficial impact of an increased surface roughness [11-15], we assume that our findings will facilitate to tailor the functional properties of carbon fibers as reinforcement material for polymers and concrete aiming to increase the fiber-matrix bonding strength. The demonstrated optical properties of LIPSS on fibers might be of potential interest concerning plasmonic-enhanced light coupling into fiber materials.

*Acknowledgements*

The SEM facilities of the Jena Center for Soft Matter (JCSM) were established with a grant from the German Research Council (DFG) and the European Fonds for Regional Development (EFRE). We thank Stephanie Höppener and Ulrich S. Schubert for access to Raman spectroscopy measurements. Furthermore, we thankfully acknowledge Michael Kracker and Hannes Engelhardt for their support during the tensile strength measurements. Financial support of the DFG trough research grant TU149/5-1 and research infrastructure grant INST 275/257-1 FUGG is acknowledged. This work has also received funding from the European Union's Horizon 2020 research and innovation program under grant agreement No 696656 (Graphene Flagship).

*References*

[1]    Morgan P. Carbon Fibers and Their Composites. Florida: CRC Press; 2005.

[2]    Chung DDL. Carbon Fiber Composites. Boston: Butterworth-Heinemann; 1994.

[3]    Mechtcherine V. Novel cement-based composites for the strengthening and repair of concrete structures. Constr Build Mater. 2013;41:365-73.

[4]   Schneider K, Lieboldt M, Liebscher M, Fröhlich M, Hempel S, Butler M, et al. Mineral-Based Coating of Plasma-Treated Carbon Fibre Rovings for Carbon Concrete Composites with Enhanced Mechanical Performance. Materials. 2017;10(4):360.

[5]   Barnett FR, Norr MK. A three-dimensional structural model for a high modulus pan-based carbon fibre. Composites. 1976;7(2):93-9.

[6]   Huang XS. Fabrication and Properties of Carbon Fibers. Materials. 2009;2(4):2369-403.

[7]   Zhang RL, Gao B, Ma QH, Zhang J, Cui HZ, Liu L. Directly grafting graphene oxide onto carbon fiber and the effect on the mechanical properties of carbon fiber composites. Materials & Design. 2016;93:364-9.

[8]   Sharma SP, Lakkad SC. Effect of CNTs growth on carbon fibers on the tensile strength of CNTs grown carbon fiber-reinforced polymer matrix composites. Composites Part A: Appl Sci Manuf. 2011;42(1):8-15.

[9]   Lv P, Feng YY, Zhang P, Chen HM, Zhao NQ, Feng W. Increasing the interfacial strength in carbon fiber/epoxy composites by controlling the orientation and length of carbon nanotubes grown on the fibers. Carbon. 2011;49(14):4665-73.

[10]  Servinis L, Henderson LC, Andrighetto LM, Huson MG, Gengenbach TR, Fox BL. A novel approach to functionalise pristine unsized carbon fibre using in situ generated diazonium species to enhance interfacial shear strength. J Mater Chem A. 2015;3(7):3360-71.

[11]  Oliveira V, Sharma SP, de Moura MFSF, Moreira RDF, Vilar R. Surface treatment of CFRP composites using femtosecond laser radiation. Opt Las Engin. 2017;94:37-43.

[12]  Drzal LT, Sugiura N, Hook D. The role of chemical bonding and surface topography in adhesion between carbon fibers and epoxy matrices. Compos Interfaces. 1996;4(5):337-54.

[13]  Tiwari S, Bijwe J. Surface Treatment of Carbon Fibers - A Review. 2nd International Conference on Innovations in Automation and Mechatronics Engineering, Iciame 2014. 2014;14:505-12.

[14]  Wu S, Liu YQ, Ge YC, Ran LP, Peng K, Yi MZ. Surface structures of PAN-based carbon fibers and their influences on the interface formation and mechanical properties of carbon-carbon composites. Composites Part A: Appl Sci Manuf. 2016;90:480-8.

[15]  Wan YZ, Wang YL, Li QY, Dong XH. Influence of surface treatment of carbon fibers on interfacial adhesion strength and mechanical properties of PLA-based composites. J Appl Polym Sci. 2001;80(3):367-76.

[16] Bonse J, Kirner SV, Höhm S, Epperlein N, Spaltmann D, Rosenfeld A, et al. Applications of laser-induced periodic surface structures (LIPSS). Proc SPIE. 2017;10092:100920N.

[17] Bonse J, Krüger J, Höhm S, Rosenfeld A. Femtosecond laser-induced periodic surface structures. J Laser Appl. 2012;24(4):042006.

[18] Müller F, Kunz C, Gräf S. Bio-Inspired Functional Surfaces Based on Laser-Induced Periodic Surface Structures. Materials. 2016;9(6):476.

[19] Bonse J, Höhm S, Kirner SV, Rosenfeld A, Krüger J. Laser-Induced Periodic Surface Structures-A Scientific Evergreen. IEEE J Select Top Quant Electron. 2017;23(3):9000615.

[20] Hermens U, Kirner SV, Emonts C, Comanns P, Skoulas E, Mimidis A, et al. Mimicking lizard-like surface structures upon ultrashort laser pulse irradiation of inorganic materials. Appl Surf Sci. 2017;418:499-507.

[21] Gräf S, Kunz C, Müller F. Formation and Properties of Laser-Induced Periodic Surface Structures on Different Glasses. Materials. 2017;10(8):933.

[22] Sipe JE, Young JF, Preston JS, van Driel HM. Laser-induced periodic surface structure. I. Theory. Phys Rev B. 1983;27(2):1141-54.

[23] Bonse J, Rosenfeld A, Krüger J. On the role of surface plasmon polaritons in the formation of laser-induced periodic surface structures upon irradiation of silicon by femtosecond-laser pulses. J Appl Phys. 2009;106(10):104910.

[24] Huang M, Zhao FL, Cheng Y, Xu NS, Xu ZZ. Origin of Laser-Induced Near-Subwavelength Ripples: Interference between Surface Plasmons and Incident Laser. Acs Nano. 2009;3(12):4062-70.

[25] Gregorcic P, Sedlacek M, Podgornik B, Reif J. Formation of laser-induced periodic surface structures (LIPSS) on tool steel by multiple picosecond laser pulses of different polarizations. Appl Surf Sci. 2016;387:698-706.

[26] Reif J, Costache F, Henyk M, Pandelov SV. Ripples revisited: non-classical morphology at the bottom of femtosecond laser ablation craters in transparent dielectrics. Appl Surf Sci. 2002;197:891-5.

[27] Li X-F, Zhang C-Y, Li H, Dai Q-F, Lan S, Tie S-L. Formation of 100-nm periodic structures on a titanium surface by exploiting the oxidation and third harmonic generation induced by femtosecond laser pulses. Opt Express. 2014;22(23):28086-99.

[28] Borowiec A, Haugen HK. Subwavelength ripple formation on the surfaces of compound semiconductors irradiated with femtosecond laser pulses. Appl Phys Lett. 2003;82(25):4462-4.

[29] Feng P, Zhang N, Wu H, Zhu X. Effect of ambient air on femtosecond laser ablation of highly oriented pyrolytic graphite. Opt Lett. 2015;40(1):17-20.

[30] Huang M, Zhao F, Cheng Y, Xu N, Xu Z. Large area uniform nanostructures fabricated by direct femtosecond laser ablation. Opt Express. 2008;16(23):19354-65.

[31] Stępak B, Dzienny P, Franke V, Kunicki P, Gotszalk T, Antończak A. Femtosecond laser-induced ripple patterns for homogenous nanostructuring of pyrolytic carbon heart valve implant. Appl Surf Sci. 2018;436:682-9.

[32] Golosov EV, Ionin AA, Kolobov YR, Kudryashov SI, Ligachev AE, Makarov SV, et al. Near-threshold femtosecond laser fabrication of one-dimensional subwavelength nanogratings on a graphite surface. Phys Rev B. 2011;83(11).

[33] Sajzew R, Schröder J, Kunz C, Engel S, Müller FA, Gräf S. Femtosecond laser-induced surface structures on carbon fibers. Opt Lett. 2015;40(24):5734-7.

[34] Nathala CSR, Ajami A, Ionin AA, Kudryashov SI, Makarov SV, Ganz T, et al. Experimental study of fs-laser induced sub-100-nm periodic surface structures on titanium. Opt Express. 2015;23(5):5915-29.

[35] Höhm S, Rosenfeld A, Krüger J, Bonse J. Femtosecond laser-induced periodic surface structures on silica. J Appl Phys. 2012;112(1):014901.

[36] Calvani P, Bellucci A, Girolami M, Orlando S, Valentini V, Lettino A, et al. Optical properties of femtosecond laser-treated diamond. Appl Phys A Mater Sci Process. 2014;117(1):25-9.

[37] Vorobyev AY, Guo C. Colorizing metals with femtosecond laser pulses. Appl Phys Lett. 2008;92(4):041914.

[38] Ahsan MS, Ahmed F, Kim YG, Lee MS, Jun MBG. Colorizing stainless steel surface by femtosecond laser induced micro/nano-structures. Appl Surf Sci. 2011;257(17):7771-7.

[39] Ou ZG, Huang M, Zhao FL. Colorizing pure copper surface by ultrafast laser-induced near-subwavelength ripples. Opt Express. 2014;22(14):17254-65.

[40] Dusser B, Sagan Z, Soder H, Faure N, Colombier JP, Jourlin M, et al. Controlled nanostructrures formation by ultra fast laser pulses for color marking. Opt Express. 2010;18(3):2913-24.

[41] Veiko V, Karlagina Y, Moskvin M, Mikhailovskii V, Odintsova G, Olshin P, et al. Metal surface coloration by oxide periodic structures formed with nanosecond laser pulses. Opt Las Engin. 2017;96:63-7.

[42] Kuladeep R, Dar MH, Deepak KLN, Rao DN. Ultrafast laser induced periodic sub-wavelength aluminum surface structures and nanoparticles in air and liquids. J Appl Phys. 2014;116(11):113107.

[43] Orazi L, Gnilitskyi I, Serro AP. Laser Nanopatterning for Wettability Applications. J Micro Nano-Manuf. 2017;5(2):021008.

[44] Obona JV, Ocelik V, Skolski JZP, Mitko VS, Romer GRBE, in't Veld AJH, et al. On the surface topography of ultrashort laser pulse treated steel surfaces. Appl Surf Sci. 2011;258(4):1555-60.

[45] Huerta-Murillo D, Aguilar-Morales AI, Alamri S, Cardoso JT, Jagdheesh R, Lasagni AF, et al. Fabrication of multi-scale periodic surface structures on Ti-6Al-4V by direct laser writing and direct laser interference patterning for modified wettability applications. Opt Las Engin. 2017;98:134-42.

[46] Pan A, Si J, Chen T, Li C, Hou X. Fabrication of two-dimensional periodic structures on silicon after scanning irradiation with femtosecond laser multi-beams. Appl Surf Sci. 2016;368:443-8.

[47] Huang Y, Young RJ. Microstructure and mechanical properties of pitch-based carbon fibres. J Mater Sci. 1994;29(15):4027-36.

[48] Kaburagi Y, Yoshida A, Hishiyama Y. Chapter 7 - Raman Spectroscopy A2 - Inagaki, Michio. In: Kang F, ed. Materials Science and Engineering of Carbon: Butterworth-Heinemann 2016, p. 125-52.

[49] Ferrari AC, Basko DM. Raman spectroscopy as a versatile tool for studying the properties of graphene. Nature Nanotechnology. 2013;8(4):235-46.

[50] Ferrari AC. Raman spectroscopy of graphene and graphite: Disorder, electron–phonon coupling, doping and nonadiabatic effects. Solid State Commun. 2007;143(1):47-57.

[51] Bonse J, Hertwig A, Koter R, Weise M, Beck U, Reinstadt P, et al. Femtosecond laser pulse irradiation effects on thin hydrogenated amorphous carbon layers. Appl Phys A Mater Sci Process. 2013;112(1):9-14.

[52] Gries WH. A universal predictive equation for the inelastic mean free pathlengths of x-ray photoelectrons and Auger electrons. Surf Interface Anal. 1996;24(1):38-50.

[53] Santiago F, Mansour AN, Lee RN. XPS Study of Sizing Removal from Carbon-Fibers. Surf Interface Anal. 1987;10(1):17-22.

[54] Dilsiz N, Wightman JP. Surface analysis of unsized and sized carbon fibers. Carbon. 1999;37(7):1105-14.

[55] Zhang RL, Zhang JS, Zhao LH, Sun YL. Sizing Agent on the Carbon Fibers Surface and Interface Properties of Its Composites. Fibers and Polymers. 2015;16(3):657-63.

[56] Bonse J, Munz M, Sturm H. Structure formation on the surface of indium phosphide irradiated by femtosecond laser pulses. J Appl Phys. 2005;97(1):013538.

[57] Freitag C, Weber R, Graf T. Polarization dependence of laser interaction with carbon fibers and CFRP. Opt Express. 2014;22(2):1474-9.

[58] Djurisic AB, Li EH. Optical properties of graphite. J Appl Phys. 1999;85(10):7404-10.

[59] Sokolowski-Tinten K, von der Linde D. Generation of dense electron-hole plasmas in silicon. Phys Rev B. 2000;61(4):2643-50.

[60] Raether H. Surface-Plasmons on Smooth and Rough Surfaces and on Gratings. Springer Tracts in Modern Physics. 1988;111:1-133.

[61] Derrien TJY, Krüger J, Bonse J. Properties of surface plasmon polaritons on lossy materials: lifetimes, periods and excitation conditions. J Opt. 2016;18(11):115007.

[62] Skolski JZP, Romer GRBE, Obona JV, Ocelik V, in 't Veld AJH, De Hosson JTM. Laser-induced periodic surface structures: Fingerprints of light localization. Phys Rev B. 2012;85(7):075320.

[63] Rudenko A, Colombier JP, Höhm S, Rosenfeld A, Krüger J, Bonse J, et al. Spontaneous periodic ordering on the surface and in the bulk of dielectrics irradiated by ultrafast laser: a shared electromagnetic origin. Sci Rep. 2017;7:12306.